\documentclass[conference]{IEEEtran}
\IEEEoverridecommandlockouts
\usepackage{cite}
\usepackage{amsmath,amssymb,amsfonts}
\usepackage{algorithmic}
\usepackage{tabularx}
\usepackage{graphicx}
\usepackage{textcomp}
\usepackage{xcolor}
\usepackage{microtype} 
\def\BibTeX{{\rm B\kern-.05em{\sc i\kern-.025em b}\kern-.08em
    T\kern-.1667em\lower.7ex\hbox{E}\kern-.125emX}}

\begin{document}

\title{Computer Vision Intelligence Test Modeling and Generation: A Case Study on Smart OCR}

\author{\IEEEauthorblockN{Jing Shu, Bing-Jiun Miu}
\IEEEauthorblockA{\textit{Department of Computer Engineering} \\
\textit{San Jose State University}\\
San Jose, USA \\
\{jing.shu, bing-jiun.miu\}@sjsu.edu}
\and
\IEEEauthorblockN{Eugene Chang}
\IEEEauthorblockA{\textit{ALPSTouchStone, Inc.} \\
USA \\
eugenechang2002@gmail.com}
\and
\IEEEauthorblockN{Jerry Gao$^{\star}$, Jun Liu$^{\star}$}
\IEEEauthorblockA{\textit{Department of Computer Engineering} \\
\textit{San Jose State University}\\
San Jose, USA \\
\{jerry.gao, junliu\}@sjsu.edu}
\thanks{$*$: \emph{Corresponding authors.}}
}

\maketitle

\begin{abstract}
AI-based systems possess distinctive characteristics and introduce challenges in quality evaluation at the same time. Consequently, ensuring and validating AI software quality is of critical importance. In this paper, we present an effective AI software functional testing model to address this challenge. Specifically, we first present a comprehensive literature review of previous work, covering key facets of AI software testing processes. We then introduce a 3D classification model to systematically evaluate the image-based text extraction AI function, as well as test coverage criteria and complexity. To evaluate the performance of our proposed AI software quality test, we propose four evaluation metrics to cover different aspects. Finally, based on the proposed framework and defined metrics, a mobile Optical Character Recognition (OCR) case study is presented to demonstrate the framework’s effectiveness and capability in assessing AI function quality.
\end{abstract}

\begin{IEEEkeywords}
Artificial Intelligence, AI software testing, computer vision, optical character recognition, quality assurance
\end{IEEEkeywords}

\section{Introduction}\label{A}
Artificial Intelligence (AI) has experienced significant growth and evolution in the past few years \cite{b1}.
Integrating AI functions, such as identification, recognition, prediction, and recommendation, is increasingly prevalent in software systems.
As a result, testing AI software, which refers to diverse quality testing activities for AI-based software systems \cite{b2}, is essential to guarantee that AI features operate correctly, reliably, and consistently.
However, AI software testing differs from conventional software testing since AI features have unique characteristics like large-scale unstructured input data, unpredicted scenarios, uncertainty in system outputs, responses or actions, and data-driven learning features \cite{b3}. This introduces challenges in AI software testing, such as the interpretability of ML and DL models, lack of clear specifications and defined requirements, test input generation, defining test oracles, and managing dynamic environments \cite{b4}. To address these issues, extensive research has been done on a variety of AI testing topics, including test modeling \cite{b5}\cite{b6}\cite{b7}\cite{b8}, test case generation \cite{b9}\cite{b10}\cite{b11}, test framework, and automation \cite{b8}\cite{b9}\cite{b12}\cite{b13}. AI software systems, based on their functionalities, can be classified into categories like expert systems, computer vision, speech recognition, NLP, and business intelligence systems. Given that different AI systems possess distinct characteristics, they need varied testing approaches for validation.

Optical Character Recognition (OCR) is a fundamental computer vision technique for identifying text in images or scanned documents \cite{b14}.
In modern OCR systems, text recognition usually consists of 6 steps: image acquisition, pre-processing, segmentation, feature extraction, classification, and post-processing \cite{b15}. Techniques are utilized in every step to improve the overall accuracy of the OCR system, such as the Conventional Neural Network (CNN), one of the most popular techniques employed for OCR \cite{b17}\cite{b18}.
Stephen V. Rice introduced a series of performance measures, including character accuracy, word accuracy, non-stopword accuracy, and phrase accuracy for the page-reading system in his 1996 doctoral dissertation \cite{b21}. Character Error Rate (CER) and Word Error Rate (WER), which are the inverted accuracy, have also been commonly used in OCR evaluation. When documents have multiple columns or distinct text blocks, layout analysis determines the text segmentations, reading order, and geometric position, which becomes a crucial aspect of the evaluation process. Modified CER and WER with configurable penalties in reading order, over-segmentation or under-segmentation of text lines are proposed for end-to-end evaluation \cite{b22}. Based on text bounding coordinates, Hwang \textit{et al.} \cite{b23} defined DISGO WER to evaluate four types of errors: Deletion, Insertion, Substitution, and Grouping/Ordering errors. In some situations, the text reading order is not critical, which means different reading orders are allowable. In this case, a flexible character accuracy independent of the reading order is proposed by Clausner \textit{et al.} \cite{b24}. These metrics are used to evaluate OCR systems’ performance based on various datasets, such as IMPACT \cite{b25}, which has over 600,000 representative document images collected from major European libraries, and COCO-Text \cite{b26}, which contains text in natural images. Although researchers and practitioners could assess the overall performance of OCR algorithms and systems or a specific method like parameter tuning from these metrics, a systematic way to identify the bottlenecks is still needed.

Software testing generally consists of unit testing, integration testing, function testing, and system testing.
As pointed out by Gao \textit{et al.} (2021) \cite{b5}, the process of testing AI functions primarily comprises test modeling, test case generation, test execution, and test quality evaluation.

\textit{1) Test Modeling}: Several typical validation approaches for traditional software have been applied for AI software testing, like Metamorphic Testing (MT). MT, a methodology initially introduced by Chen \textit{et al.} (1998) \cite{b6}, employs the properties of functions such that it is possible to enable the prediction of specific output changes in response to particular modifications in the input. In this way, it could address “Test Oracle Problems”.
C. Tao \textit{et al.} (2019) \cite{b7} propose an MT-based testing model, especially for facial age recognition systems. Y. He \textit{et al.} (2023) \cite{b8} propose a semantic tree model for generating driving scenarios in Autonomous Vehicle (AV) testing, along with a 3D test model for performance evaluation through automation tests. This 3D test model includes a context semantic tree detailing constant aspects through a test case and separating from the dynamic agents, an input semantic tree modeling dynamic agents like pedestrians, vehicles, and major landmarks, and an output semantic tree capturing behaviors with the given context and input. Based on this model, complete scenarios are generated.

\textit{2) Test Case Generation}: The distinctive features of AI software, such as large-scale input data and data-driven learning features, make it extremely difficult and expensive to generate test cases and validate results. To tackle these challenges, H. Zhu \textit{et al.} (2019) \cite{b9} present a novel approach called datamorphic testing for AI applications. In the testing, datamorphism involves transforming existing test data, termed seed, to create new test data, called mutants. These alterations result in a variety of mutants derived from the original seed data. D. Berend (2021) \cite{b10} points out that in recent studies, the distribution of generated test data has not been considered when designing testing techniques for AI software. Therefore, the author proposes an innovative distribution-aware testing methodology that seeks to generate new unseen test cases related to the fundamental system task. J. Huang \textit{et al.} (2022) \cite{b11} have found that 44\% of the test cases generated by the state-of-the-art test approaches are false alarms because of inconsistent and unnatural issues. To address this problem, they propose AEON for Automatic Evaluation of NLP test.

\textit{3) Test Framework and Automation}: To speed up the AI-based system validation process and reduce its cost and time, numerous studies have focused on innovative testing frameworks and test automation. L. Li \textit{et al.} (2023) \cite{b12} propose an innovative AI model evaluation framework based on a quality evaluation model. This model is distinguished by three attributes: mathematical, comprehensive, and software, each encompassing various quality dimensions. H. Zhu \textit{et al.} (2020) \cite{b13} extend the datamorphic testing framework by categorizing software artifacts into two types: entities and morphisms. They also present an automated testing tool named Morphy.

The rest of this paper is organized as follows.
Section \ref{C} describes the proposed AI test modeling and presents the test coverage criteria and complexity. The proposed evaluation metrics are introduced in Section \ref{D}. The experimental results and analysis are discussed in Section \ref{E}. The conclusion and future work are given in Section \ref{F}.

\section{AI Function Test Modeling}\label{C}
\subsection{Preliminaries: OCR Evaluation}
The most common way to evaluate OCR performance is to compare OCR result text with Ground Truth (GT). The edit distance between OCR text and GT is typically used to define OCR accuracy. Levenshtein Distance \cite{b27} is one type of edit distance, in which each insertion, deletion, and substitution corresponds to one edit operation. Let R = $r_1r_2r_3\cdots r_m$ be an OCR result text string of m symbols and G = $g_1g_2g_3 \cdots g_n$ be the correct text of n symbols. The Character Accuracy (CA) \cite{b21} is defined as $\text{CA}= \frac{n-E}{n},$
where the number of errors $E$ is the edit distance that denotes the minimum number of edit operations of transferring string R into string G.
Similarly, Word Accuracy is the percentage of words that are correctly recognized. Character Accuracy indicates how well the OCR system recognizes characters. But in some cases, not all the characters are interested. Therefore, Non-Stop word Accuracy, Phrase Accuracy (accuracy over a sequence of k characters), and Accuracy by Character Class (accuracy over a subset of characters) are also introduced.

To calculate the edit distance, both the serialized OCR output and the serialized GT text are required. The complexity of a text’s layout can impact OCR accuracy, especially due to the reading order of various text blocks or columns. However, in some cases, the reading order is less critical.
Clausner \textit{et al.} \cite{b24} proposed flexible character accuracy to solve this problem. The texts to be compared are broken down into smaller segments. Then, individual edit distance calculations will be conducted on these segments. Finally, these distances are aggregated to obtain an overall measure of character accuracy.
Layout analysis \cite{b28} is another important step if the OCR system aims to extract the geometric and logical structure of the text in an image or scanned document.

According to the related work, there are different AI software testing approaches, such as metamorphic testing and model-based testing. Unlike NLP, the OCR text extraction AI feature has a predictable output for a specific image or scanned document. This predictability stems from the fact that for a specific picture or scanned document, the text content on it is objectively present. As a result, the OCR process inherently has a deterministic output. Recognizing this characteristic, it becomes crucial to identify a suitable and cost-effective approach for validating the OCR function.

This paper is focused on an AI function that takes image-based receipts as input data and converts them to editable text via OCR technology. Recent receipt OCR applications mainly focus on Key Information Extraction (KIE) such as merchant name, address, and total expense, which does not fulfill our requirements. So, we resort to “OCR text scanner” mobile applications like \textit{CamScanner} and \textit{Scan Pro} for our case study.

\subsection{Context Classification Modeling}
In this paper, a 3D classification testing model \cite{b5} is designed for the image-based receipt text extraction AI function. The term “3D” refers to 3 dimensions: context, input, and output. The model is based on three types of classification tree models to detail different aspects of scenarios: context classification tree model, input classification tree model, and output classification tree model. Then, the specific conditions of each scenario, which are related to each test case, are visualized in a 3D decision table.
Meanwhile, test coverage criteria and test complexity are discussed to provide a thorough explanation of these critical aspects.

\begin{figure}[!tbp]
\centerline{\includegraphics[width=0.8\linewidth]{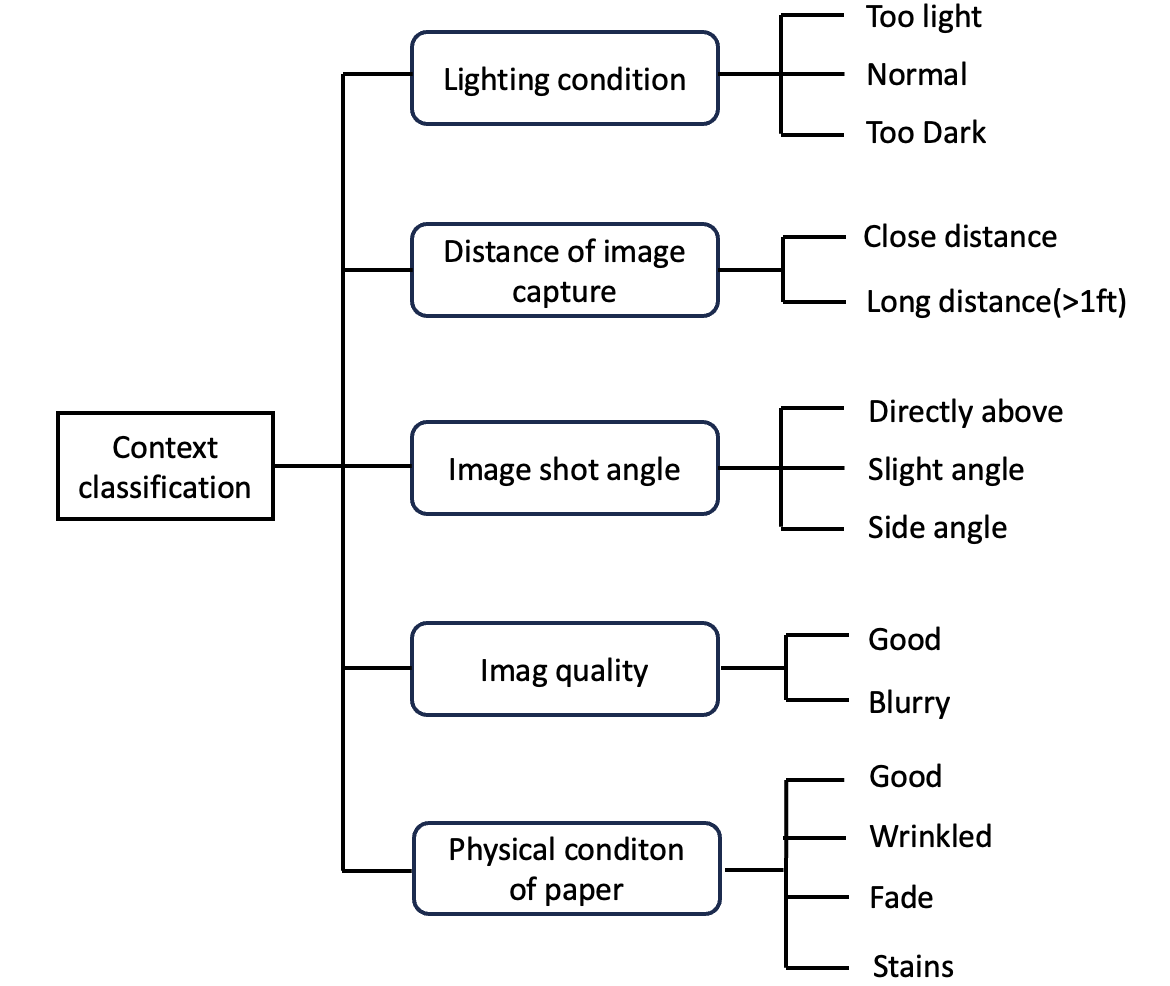}}
\caption{A context classification tree that depicts the environmental conditions when the image is captured.}
\label{fig2}
\end{figure}

The context of the selected AI function contains the environmental conditions when the image is captured. Many factors influence the function's performance, such as lighting conditions, image shot angle, and image quality.
Taking different factors that have an impact on final performance into account, a context classification tree has been created, as illustrated in Fig. \ref{fig2}.

\subsection{Input Classification Modeling}
\begin{figure}[!tbp]
\centerline{\includegraphics[width=\linewidth]{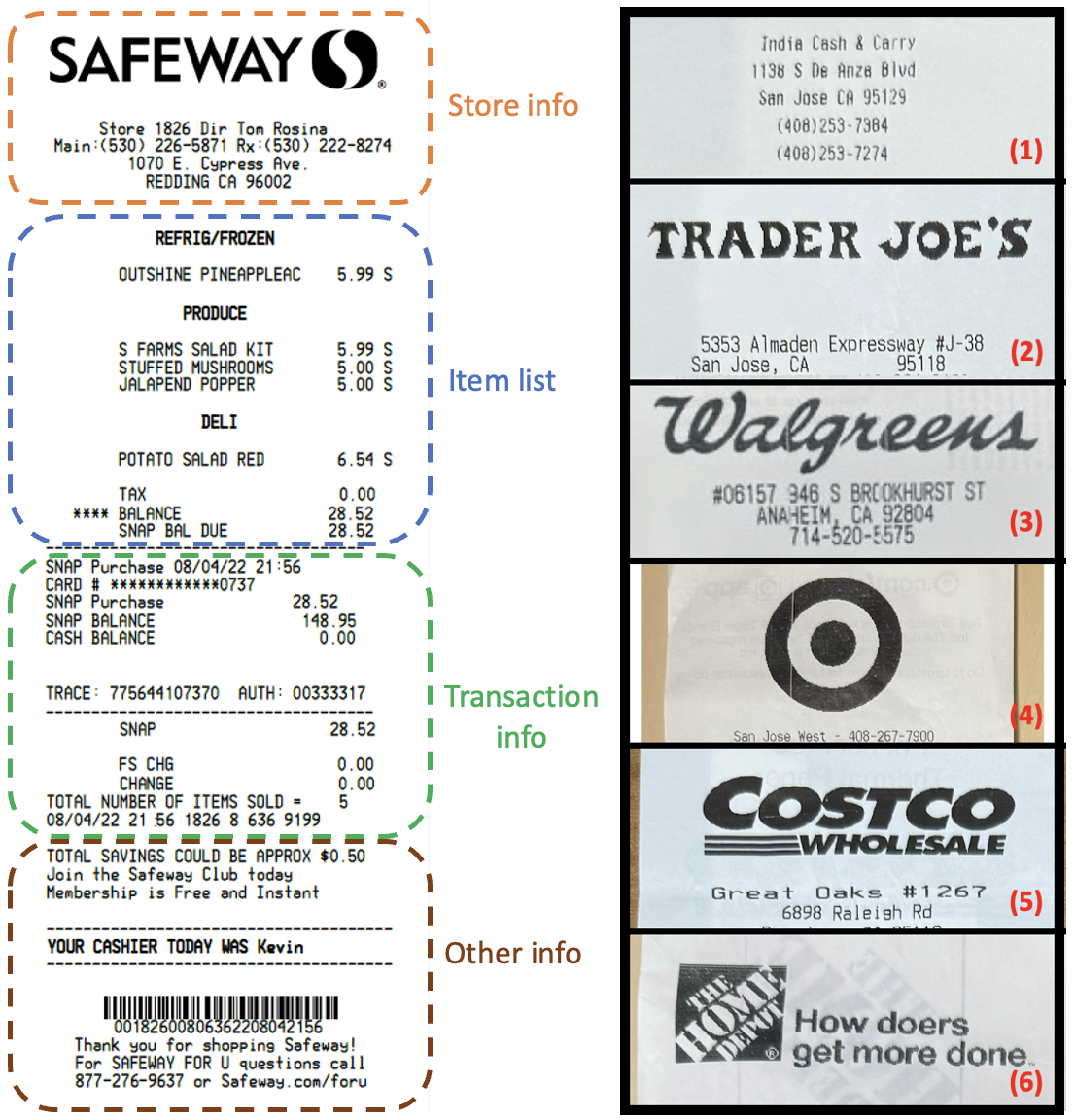}}
\caption{Left: four sections of a shopping receipt. Right: Store info in a receipt. (1) No logo; (2) Text logo – normal font; (3) Text logo – decorated font; (4) Logo w/o text; (5) Logo w/ normal font text; (6) Logo w/ decorated text.}
\label{fig3}
\end{figure}

The input aspect of the model relates to the textual contents within the image. The author divides the receipt content into four main sections: store details, item list, transaction specifics, and miscellaneous information. Consider a typical shopping receipt illustrated in Fig. \ref{fig3}. The first section of receipts typically contains the store’s name, address, and contact details, along with the store’s logo, if it exists. Logos vary widely: some are purely graphical or textual with diverse font styles, while others combine text and images. In some logos, the text is integrated into the graphic element with a colored background. The right part of Fig. \ref{fig3} shows different types of stores’ logos. The logo in (6) is harder to identify, in which text is 45 degrees rotated and embedded in a dark background.

In the purchased items section, the content is classified into three types: short, medium, and long list. This section may include discount details and special characters such as @, \#, and \%. If a credit or debit card is used for payment, the card details appear in the receipt’s transaction section. Conversely, transaction details tend to be simpler for cash payments. The miscellaneous section encompasses various other details such as barcodes, return policies, membership credits, and other information.

\begin{figure}[htbp]
\centerline{\includegraphics[width=\linewidth]{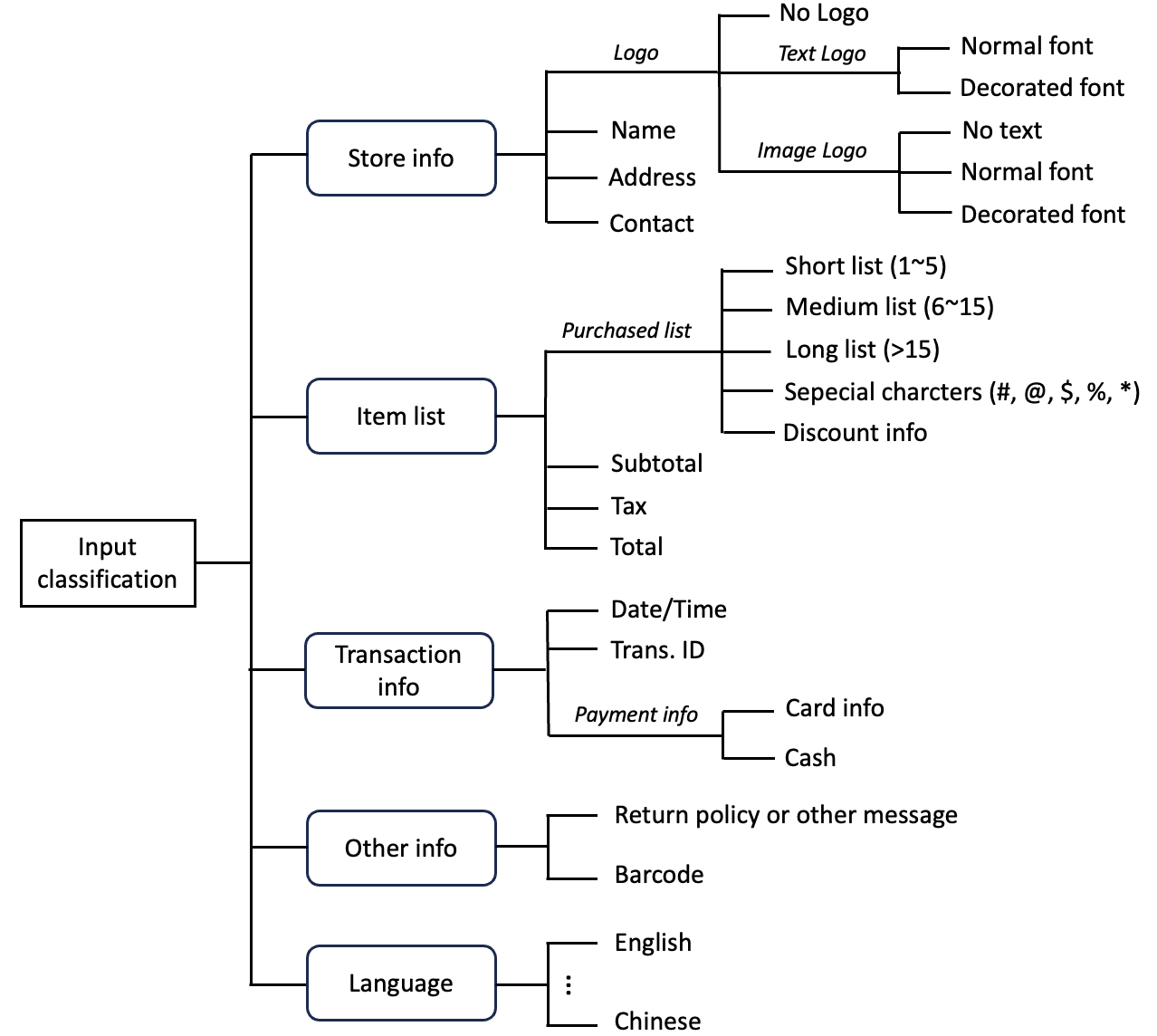}}
\caption{An input classification tree}
\label{fig4}
\end{figure}

Lastly, the language could assess the linguistic capabilities of these OCR applications under test. This involves evaluating the apps’ ability to identify various languages in receipts from different countries. Based on all these aspects, an input classification tree is built, as shown in Fig. \ref{fig4}.

\subsection{Output Classification Modeling}
According to the related work, accuracy or error rate is commonly used for OCR function evaluation. To recognize where the error comes from, total accuracy, store info section accuracy, item list section accuracy, transaction section accuracy, and miscellaneous section accuracy are defined. The method for calculating accuracy depends on the requirements. The details of evaluation metrics are discussed in Section \ref{D}.

\begin{figure}[!tbp]
\centerline{\includegraphics[width=0.75\linewidth,height=0.75\linewidth]{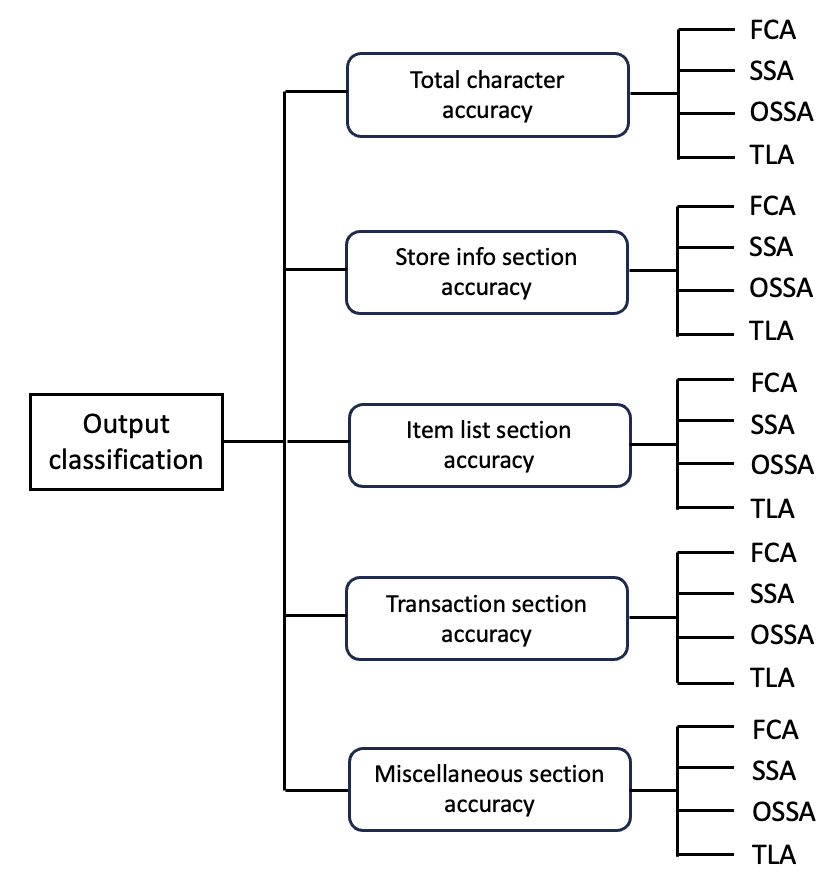}}
\caption{An output classification tree}
\label{fig5}
\end{figure}

Based on the analysis, plain text accuracy, which can be classified into four categories, is used to evaluate a test case. For the OCR output, the benchmark for a successful test (pass) could be set at a character accuracy rate of 95\% and above. This criterion ensures that the AI function not only detects text but also recognizes it with high precision. The output classification tree is shown in Fig. \ref{fig5}.

\subsection{3D Decision Table}
Based on the three classification tree models, the authors construct a 3D decision table that comprises all the considered parameters. This table presents a comprehensive view of potential test cases. Nevertheless, considering every possible combination of these parameters will lead to an extremely large number of test cases. Combinatorial testing \cite{b29} is a systematic method to select combinations of inputs, parameters, or features for testing, which has been extensively studied in the past few decades. Combinatorial T-way testing is effective in balancing the test costs and coverage. This approach can be used to generate test cases. However, the primary goal of this project is to identify the factors affecting OCR accuracy during evaluation. To isolate the influence of different factors, we change only one parameter at a time in each test to determine the most significant influencing factor in this paper. For example, if dark lighting and long-distance image capture conditions are considered simultaneously in one test and it fails, we need to identify the main contributor to the failure. Based on the test results, combinatorial testing will be carefully designed to increase test coverage later.

To reduce the test cost and find meaningful test cases, the authors choose one scenario as the base test case, including the basic control parameters as a standard. Other test cases are derived from the base test case by tweaking one context condition or input parameter at a time to collect data. This approach allows for data collection on performance and behavior changes under varying conditions.

Given these considerations, the decision table helps to select $24$ test cases, covering all the attributes of context and input classification trees. Table 1 demonstrates these selected test cases, using an $\times$ to denote the inclusion of a particular context condition or input attribute in the image and an $\circ$ to represent the expected output.
In our experiment, we use various context conditions, i.e., different lighting and distance conditions, to test the performance of our method.

\begin{table*}[htbp]
\caption{A 3D decision table encompassing all the parameters to select test cases.}
\centering{\includegraphics[width=\linewidth]{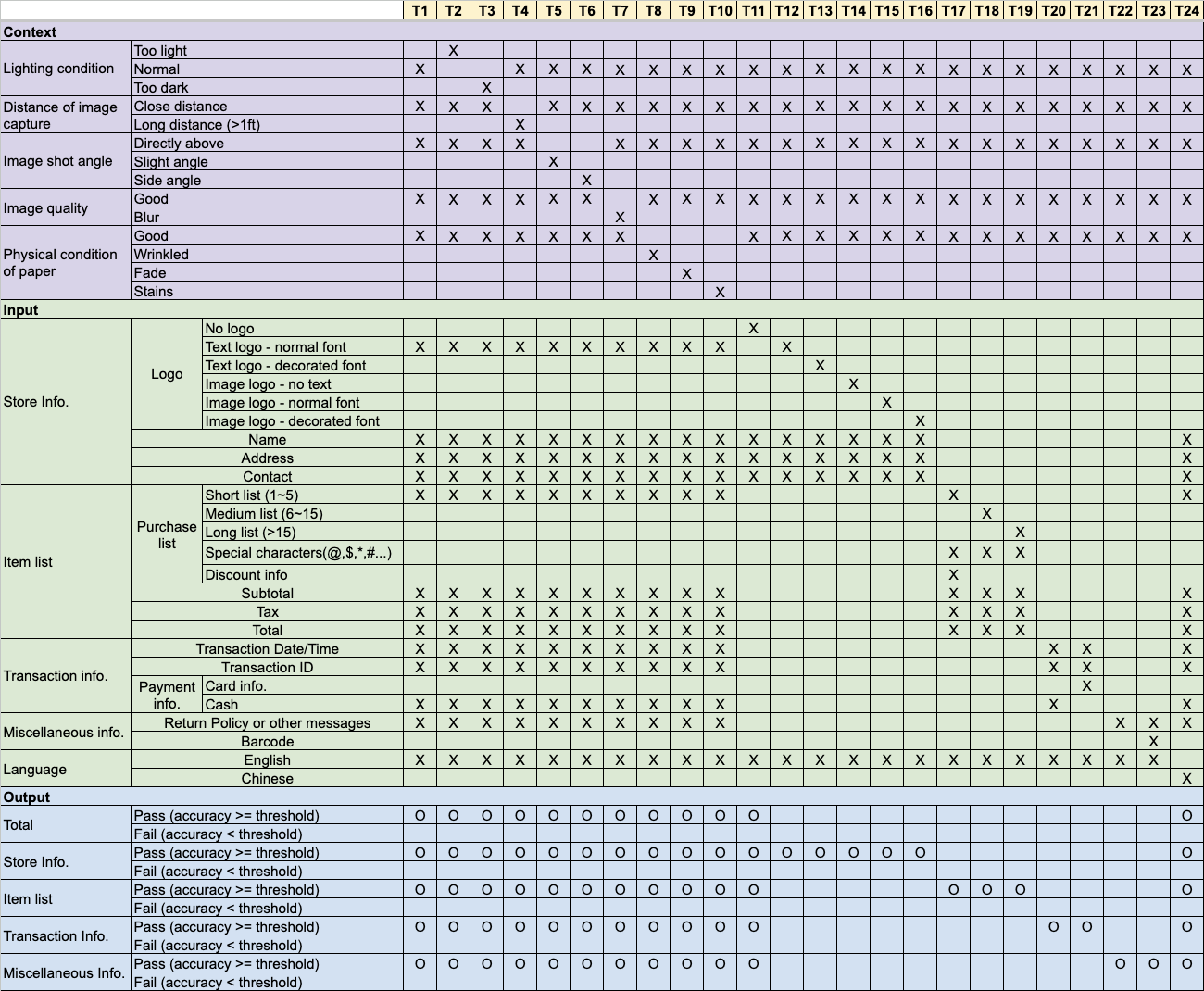}}
\label{table1}
\end{table*}

\subsection{Test Coverage Criteria}
For the given AI function, its 3D decision table has a set of context conditions, input, and output parameters. The test coverage criteria are fulfilled when any context condition, input, and output parameter are covered by at least one test case. Based on the test models, the test complexity of each classification tree can be easily computed below:
\begin{itemize}
\item Context classification test Complexity (CC) = Number of context classification stubs in Table \ref{table1} = $14$
\item Input classification test Complexity (IC) = Number of input classification stubs in Table \ref{table1} = $25$
\item Output classification test Complexity (OC) = Number of output classification stubs in Table \ref{table1} = $10$
\end{itemize}

The total test complexity is CC$\cdot$IC$\cdot$OC $= 14 \cdot 25 \cdot 10 = 3500$. However, the author reduces the test complexity to $24$ by selecting test cases based on the test coverage criteria in the decision table.

\section{Model Evaluation Metrics}\label{D}
If the sequence in which text is read doesn’t matter, which means Ground Truth is independent of reading order, a measure such as the flexible character accuracy \cite{b24} could be applied. Furthermore, if the accuracy assessment requires considering the text’s geometric position, the text position, like bounding box coordinates, should be incorporated into the error calculation process.

In this paper, text positioning is not considered in the accuracy calculation, as text scanner applications typically produce unformatted text outputs. Accuracy is focused on characters and string segments, assessed at four levels: Flex Character Accuracy (FCA), String Segment Accuracy (SSA), Ordered  String Segment Accuracy (OSSA), and Text-Line Accuracy (TLA). Characters are categorized into alphabet characters, special characters, and digits to further evaluate the OCR system’s proficiency in recognizing different types of characters. Before determining the accuracy for a specific character type, both the ground truth and OCR-generated text are preprocessed to remove other types of characters. Subsequently, the FCA algorithm and its associated formulas are utilized to calculate the accuracy of a particular character category, such as special character accuracy.

\begin{figure}[!tbp]
\centerline{\includegraphics[width=\linewidth]{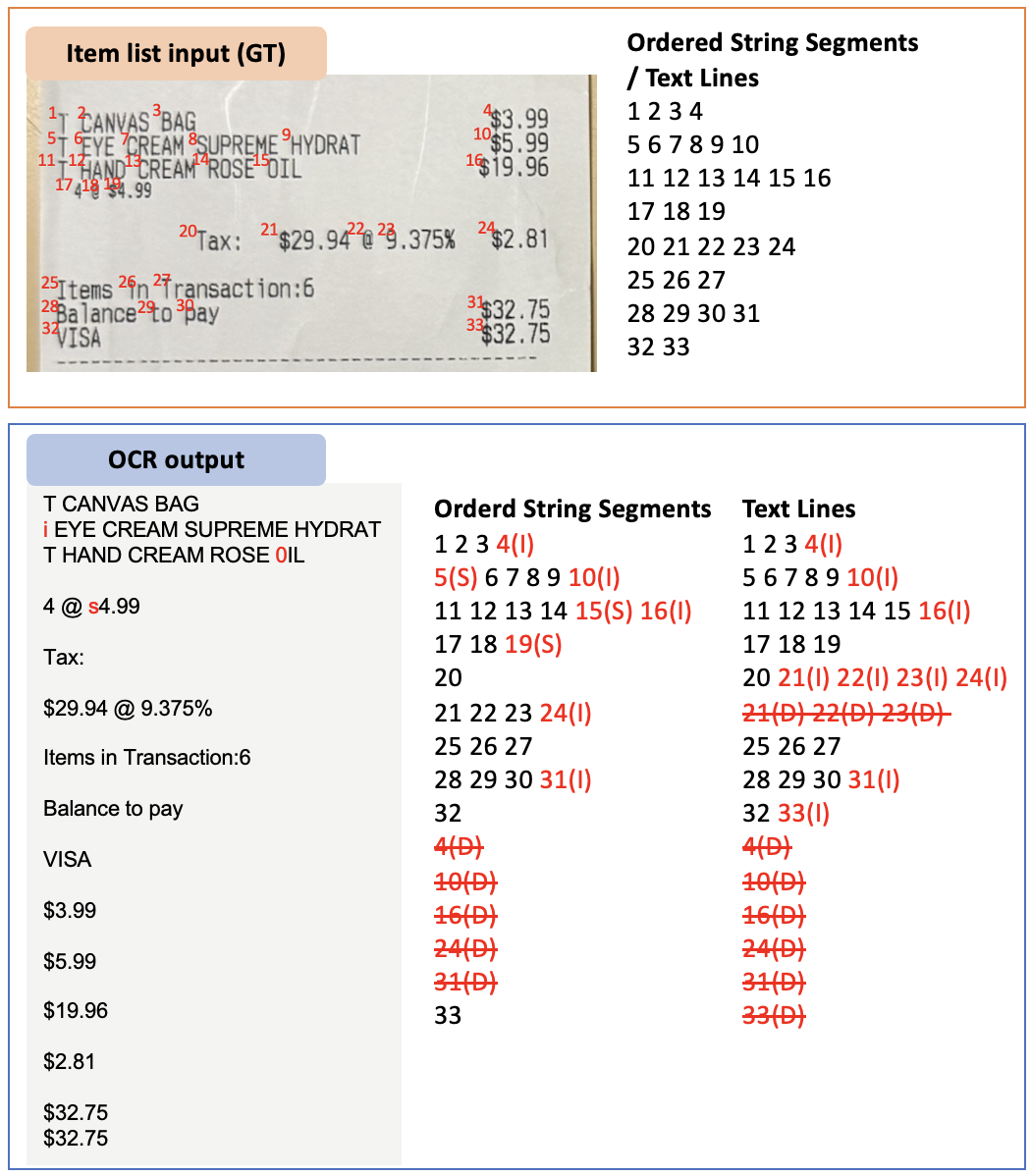}}
\caption{An example of an item list on a receipt. (1) Top part: the ground truth for this item list with each string segment sequentially numbered in red; (2) Bottom part: the OCR output of the item list with the adjusted string sequence and the allocation of items per line. Errors are indicated in red (deletions (D) crossed out, substitution: (S), insertion: (I)).}
\label{fig7}
\end{figure}

A string segment refers to a string of characters without space, which could include any combination of alphabet characters, special characters, and digits. For example, the ground truth shown in Fig. \ref{fig7} contains of 33 string segments. the 27th string segment “Transaction: 6” comprises the word “Transaction”, followed by a special character “:”, and ends with the digit “6”. In contrast, the 1st and 22nd segments contain just a single alphabet character and a special character, respectively. Before calculating SSA, every string segment is transformed into a distinct character. SSA is then calculated using the same methodology applied for FCA.

Ordered string segments adhere to a human-like reading order for receipts, from left to right and top to bottom. Like calculating SSA, the preprocessing phase for OSSA also involves converting each string segment into a unique character. For accuracy calculation, the method for Character Accuracy \cite{b21} measure is utilized, which considers the order of the strings.
Text-line Accuracy (TLA) focuses on the percentage of text lines correctly identified in the OCR result. Let $t$ be the number of text lines in the GT, and $r$ be the correct number of text lines in the OCR output. TLA is defined as $\text{TLA} = \frac{r}{t}$.
The summary of four metrics is presented in Table \ref{table2}, with an example of accuracy results shown in Fig. \ref{fig7}. These metrics are designed to evaluate OCR performance at different levels. FCA assesses individual characters without considering reading order. SSA and OSSA evaluate strings of characters (string segments), with SSA disregarding reading order and OSSA considering it. TLA focuses on the accuracy of entire text lines.

\begin{table*}[!tbp]
\caption{The Accuracies of Receipt OCR Evaluations using Different Metrics}
\centering
\begin{tabularx}{\textwidth}{|c|X|X|c|}
\hline
\textbf{Metric} & \centering{\textbf{Description}} & \centering{\textbf{Evaluation Focus}} & \textbf{Accuracy} \\
\hline
Flex Character Accuracy (FCA) & Refer to the algorithm in \cite{b24} based on character Levenshtein distance. & How well characters are correctly recognized without considering the reading order &  97.9\% \\
\hline
String Segment Accuracy (SSA) &After transforming each string segment into a distinct character, the SSA measure is the same as the FCA. & How well string segments are correctly identified without considering the reading order &  90.9\% \\
\hline
Ordered  String  Segment Accuracy (OSSA) & In the preprocessing phase, convert every string segment into a unique character. Then, the method for Character Accuracy \cite{b21} is applied to calculate OSSA. & How well string segments are correctly identified following the human-like reading order&  60.6\% \\
\hline
Text-Line Accuracy (TLA) & $\text{TLA} = \frac{r}{t}$, where \textit{t} is the number of text lines in the GT and \textit{r} is the correct number of text lines in the OCR output. & How well text lines are correctly recognized &  25.0\% \\
\hline
\end{tabularx}
\label{table2}
\end{table*}

\section{Experimental Results and Analysis}\label{E}
\subsection{Experimental Results}
\begin{figure}[!tbp]
\centerline{\includegraphics[width=\linewidth]{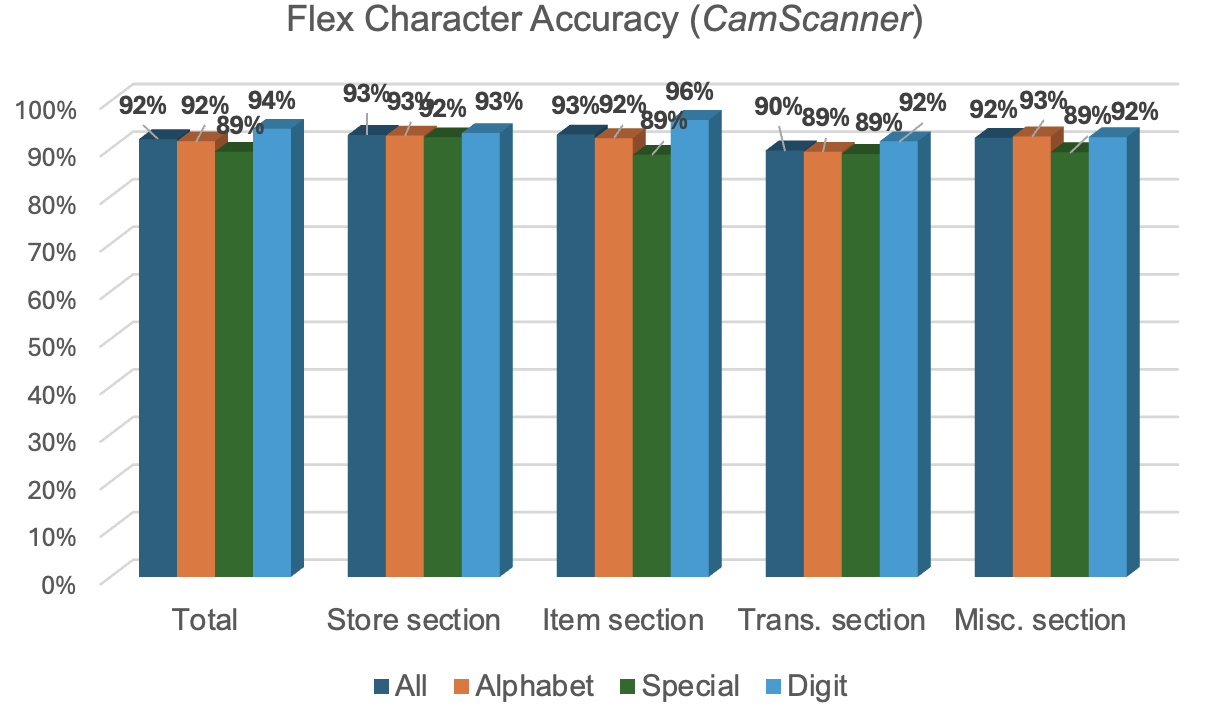}}
\caption{Different Flex Character Accuracy for \textit{CamScanner} App}
\label{fig8}
\end{figure}

\begin{figure}[!tbp]
\centerline{\includegraphics[width=\linewidth]{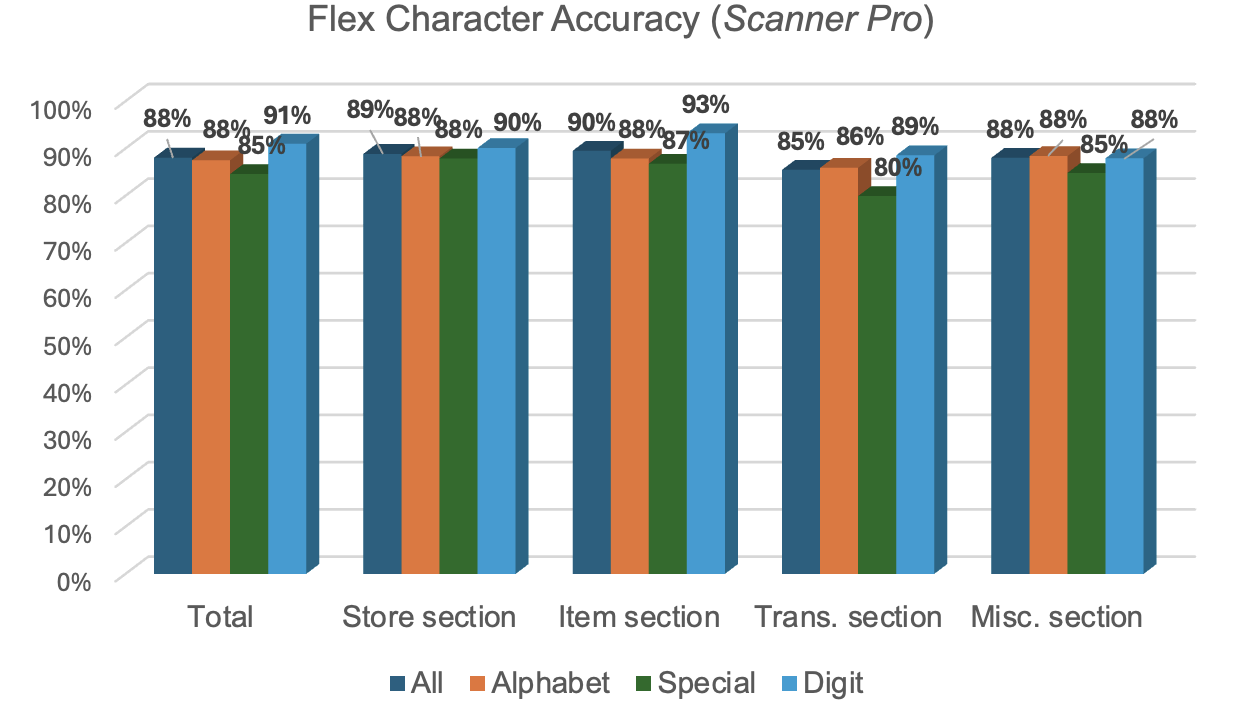}}
\caption{Different Flex Character Accuracy for \textit{Scanner Pro} App}
\label{fig9}
\end{figure}

This section presents a case study based on the proposed test model and framework for two popular “text scanner” mobile applications (\textit{CamScanner} and \textit{Scanner Pro}) to evaluate OCR AI function. The authors generate 24 test cases based on the 3D decision table. Among these test cases, pass rates based on total character accuracy (95\% and above) for \textit{CamScanner} and \textit{Scanner Pro} are 83\% and 75\%, respectively.
The accuracy results can be categorized into five sections
shown in Fig. \ref{fig8} and Fig. \ref{fig9}. \textit{CamScanner} has an overall FCA of 92\%, slightly surpassing \textit{Scanner Pro}, which has an FCA of 88\%. The transaction section typically exhibits the lowest accuracy in any application due to its complex content. Among the characters recognized, numerical digits are processed with the highest accuracy, followed by alphabet characters. Special characters tend to have the lowest accuracy rates.

\begin{figure}[!tbp]
\centerline{\includegraphics[width=0.7\linewidth]{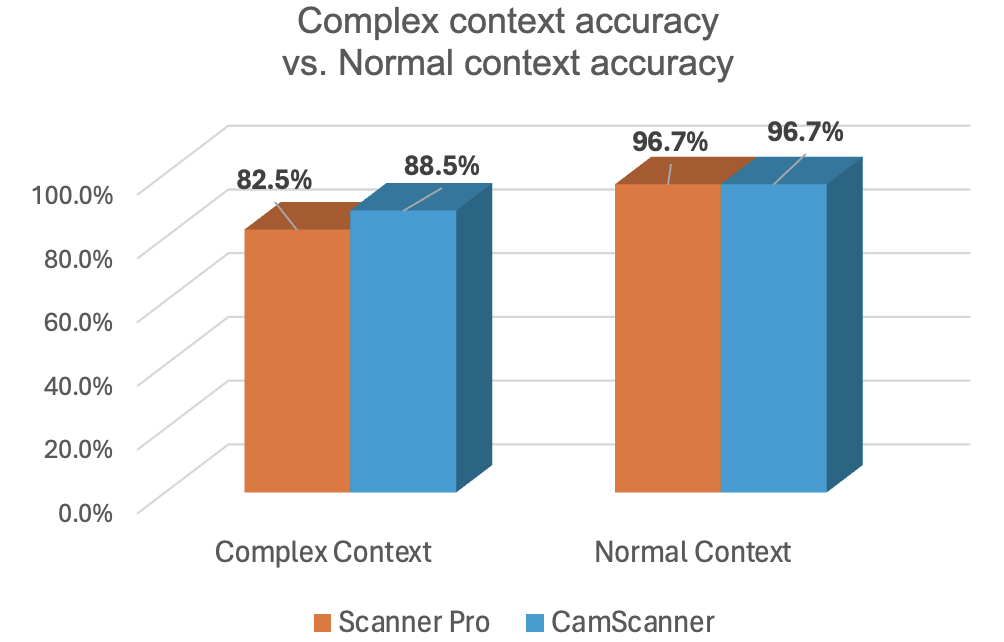}}
\caption{Comparison of complex context accuracy and normal context accuracy}
\label{fig10}
\end{figure}

Accuracy varies across test cases. By combining different test cases with complex contexts, such as poor lighting or long distances of image capture, and integrating various test cases with normal contexts, we can observe from Fig. \ref{fig10} that FCA under normal contexts can reach up to 96.7\% for both applications. However, under a complex context, \textit{CamScanner}’s accuracy falls to 88.5\%, and \textit{Scanner Pro}'s accuracy decreases to 82.5\%. \textit{CamScanner} outperforms \textit{Scanner Pro} in handling complex contexts.

\begin{figure}[!tbp]
\centerline{\includegraphics[width=\linewidth]{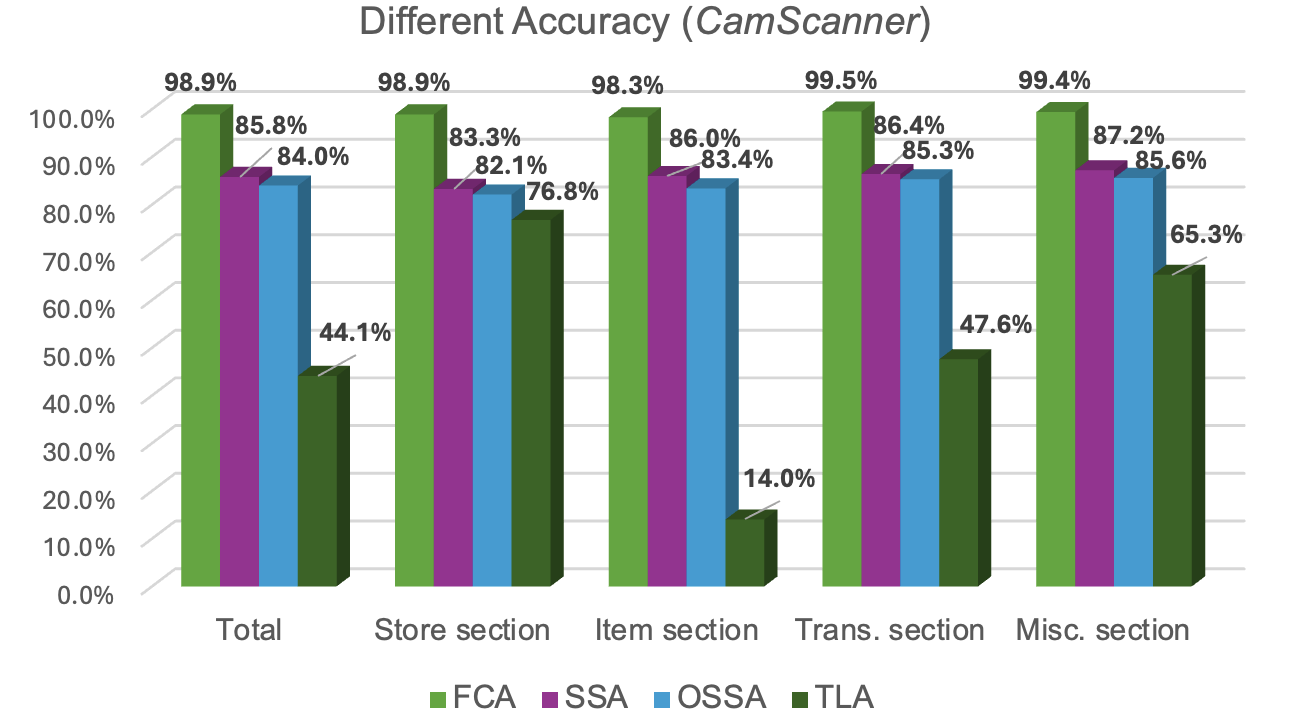}}
\caption{Different types of accuracy results for \textit{CamScanner} App}
\label{fig11}
\end{figure}

\begin{figure}[!tbp]
\centerline{\includegraphics[width=\linewidth]{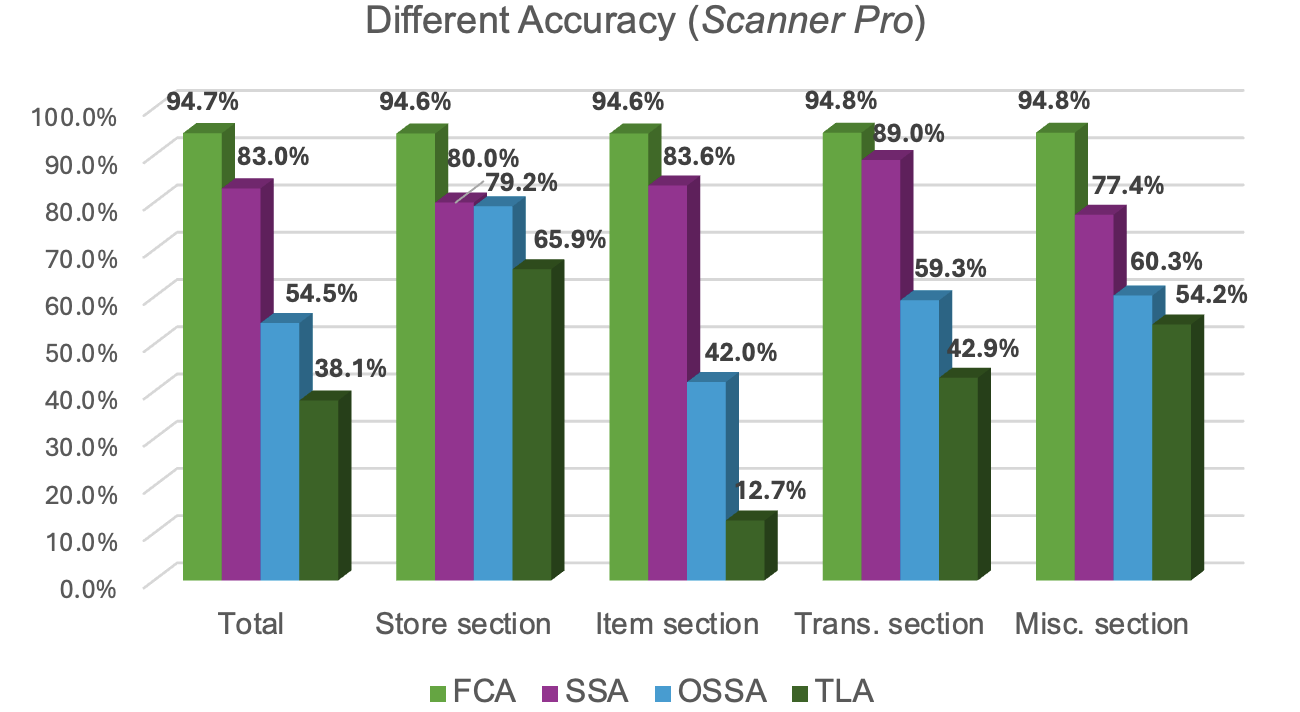}}
\caption{Different types of accuracy results for \textit{Scanner Pro} App}
\label{fig12}
\end{figure}

Exclusive of the blurry image test case with super low accuracy, accuracies are recalculated.  Fig. \ref{fig11} and Fig. \ref{fig12}. provide insights into different accuracy results for \textit{Scanner Pro} and \textit{CamScanner}. FCA emerges as the highest, with SSA ranking next. The \textit{Scanner Pro} app demonstrates a notable drop in OSSA compared to SSA, indicating a tendency for OCR outputs to rearrange words out of the original order, particularly within the item and miscellaneous sections. In contrast, the \textit{CamScanner} app maintains an SSA that is very close to OSSA, suggesting minimal alteration to word order by its OCR feature. Across both apps, TLA scores the lowest among the evaluated categories. When analyzed by section, the item section exhibits the lowest TLA, followed by the transaction section, while the store section boasts the highest TLA.

\subsection{Performance Analysis}
Based on the results obtained from our test cases, we have summarized and analyzed the strengths and weaknesses of each OCR app in the different testing scenarios. In this paper, we focus on examples of receipts. The bugs analyzed could also result in issues in other types of image-based text extraction.

\begin{figure}[!tbp]
\centerline{\includegraphics[width=0.65\linewidth,height=0.5\linewidth]{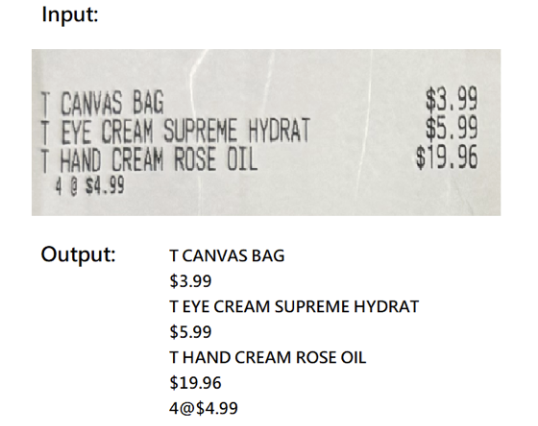}}
\caption{Line feed sample of \textit{CamScanner} }
\label{fig13}
\end{figure}

For \textit{CamScanner}, the accuracy is shown in Fig. \ref{fig11}. Its FCA is nearly perfect across all our testing scenarios.
However, there is a decline in performance for SSA and OSSA. The performance of TLA is even worse. One reason is that \textit{CamScanner} tends to ignore spaces during text detection.  This behavior might not significantly affect OCR text extraction from a human perspective, but it greatly impacts automated accuracy metrics. For example, when calculating SSA and OSSA for the case "San Jose, CA", the string segments of this line should be ['San', 'Jose,', 'CA']. However, when spaces are omitted, the content of this line in OCR output is erroneously merged into two string segments ['San', 'Jose,CA']. This leads to a reduction in both SSA and OSSA, subsequently affecting TLA. Another reason for poor performance in TLA is the line feed issue. When the string segments are spaced too far apart or are not horizontally aligned due to environmental factors, they would be fed into different lines, resulting in output that deviates from the actual result. An example of this issue is shown in Fig. \ref{fig13}. Due to the excessive spacing between string segments, the app fails to determine if they belong to the same line, leading to a discrepancy between the output number of rows and the actual results.

When processing images with good or normal contexts, the performance of the two applications is comparable. However, the OCR accuracy of \textit{Scanner Pro} typically decreases in more complex scenarios, such as images taken from a long distance or when encountering store logos with intricate fonts or image-based text. In contrast to \textit{CamScanner}, where SSA and OSSA are similar, the \textit{Scanner Pro} app often demonstrates a substantial reduction in OSSA. This is attributed to its rearrangement of string segments into a different sequence. One aspect relates to the line feed issue, which is similar to \textit{CamScanner}, is illustrated in Fig. \ref{fig7}. In this figure, we demonstrate this behavior in the \textit{Scanner Pro} app, where the spacing larger than a white space (or another specified threshold) between two string segments in the same line results in the latter being placed on a new line in the OCR output. For instance, the 21st string segment “\$29.94” is shifted to a new line, following the 20th string segment “Tax”.

\begin{figure}[!tbp]
\centerline{\includegraphics[width=0.6\linewidth,height=0.9\linewidth]{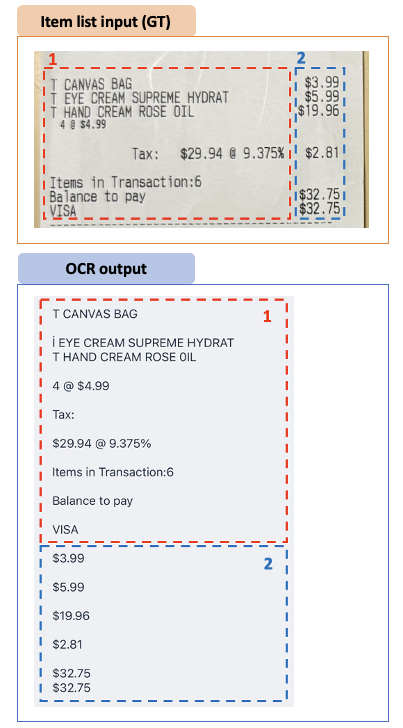}}
\caption{Line feed sample of \textit{Scanner Pro} }
\label{fig14}
\end{figure}

Moreover, excessive spacing between two string segments, like the space separating an item’s name from its price, results in the subsequent segment being misclassified as part of a different block. Fig. \ref{fig14} illustrates this issue.
The variance between SSA and OSSA serves as an indicator of this problem.

\section{Conclusion and Future Work}\label{F}
This paper discusses the challenges associated with AI software testing and provides a comprehensive literature review on related topics, including test modeling, test case generation, and test frameworks and automation. We have introduced a 3D classification testing model to systematically evaluate the image-based text extraction AI function.  Character accuracy and word accuracy are widely used in previous research.
We extend the word accuracy to string segment accuracy and utilize four specific evaluation metrics in different levels.
The effectiveness of the proposed testing model is demonstrated through an image-based text extraction case study.
Future research directions include automating this framework to increase efficiency, extending our work to other types of inputs, and making it accessible for broader applications.

\end{document}